\documentclass[a4paper,11pt]{article}
\pdfoutput=1 % if your are submitting a pdflatex (i.e. if you have
\usepackage{jinstpub} % for details on the use of the package, please see the JINST-author-manual
\usepackage{lineno}

\usepackage{amsmath,amsfonts}
\usepackage[latin1]{inputenc}
\usepackage{url}
\usepackage{graphics} %%
\usepackage{graphicx}
\usepackage{subfigure} 
\usepackage{natbib} 
\usepackage{soul} %% da rimuovere per la sottomissione
\title{Gain variations as induced by the diffuse night sky background: the ASTRI-Horn experience}
\collaboration{%
\includegraphics[height=17mm]{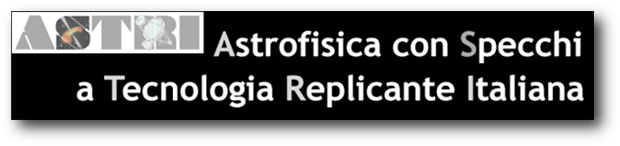}\\[6pt]
ASTRI Project collaboration}
\author[a,1]{D. Impiombato\note{Corresponding author.},}
\author[b] {A.A. Compagnino,}
\author[b]{T. Mineo,}
\author[b]{O. Catalano,}
\author[b]{S. Giarrusso,}            		
\author[b]{ M.C. Maccarone,}
\emailAdd{Domenico.Impiombato@inaf.it}
\affiliation[a]{INAF, Osservatorio Astronomico di Padova, Vicolo Osservatorio 5, 35122, Padova, Italy}
 \affiliation[b]{INAF, Istituto di Astrofisica Spaziale e Fisica Cosmica di Palermo, via U. La Malfa 153, I-90146 Palermo, Italy }
\abstract{ASTRI-Horn is the prototype of the nine telescopes that form the ASTRI Mini-Array,  under construction at the Teide Observatory in Tenerife (Spain), devoted to observe the sky in the 1--200 TeV energy band.  It adopts an innovative optical design based on a dual-mirror Schwarzschild-Couder configuration, and the camera, composed by a matrix of monolithic multipixel silicon photomultipliers (SiPMs) is managed by  ad-hoc tailored front-end electronics based on a peak-detector operation mode.
\\
During the Crab Nebula campaign in 2018-2019, ASTRI-Horn was affected by gain variations induced by high levels of night sky background. 
This paper reports the work performed to detect and quantify the effects of these gain variations in shower images. The analysis requested the use of simultaneous observations of the night sky background flux in the wavelength band 300-650 nm performed with the auxiliary instrument UVscope, a calibrated multi-anode photomultiplier  working in single counting mode.
As results, a maximum gain reduction of 15\% was observed, in agreement with the value previously computed from the  variance of the night sky background level in each image. This ASTRI-Horn gain reduction was caused by current limitation of the voltage supply.  
\\
The analysis presented in this paper provides a method to evaluate possible variations in the nominal response of SiPMs when scientific observations are performed in the presence of high night sky background as in moon conditions.}
\begin{document}
\maketitle
\flushbottom
\section{Introduction}
\label{sect:intro}

The very high-energy (VHE) $\gamma$-ray astronomy plays a crucial role in the exploration of the most violent non-thermal phenomena in the Universe. The scientific target extends from the origin of cosmic rays and dark matter to the $\gamma$-rays production in a large variety of galactic and extra-galactic sources. 
The feasibility of such studies is based on the availability of effective ground-based techniques for the detection of  $\gamma$-radiation in a very broad energy region ranging from a few tens of GeV to hundreds of TeV.
The turning point for these observations came from the Imaging Atmospheric Cherenkov Telescopes (IACTs), which observe the Cherenkov light in the air shower generated by the interaction of $\gamma$ photons with the Earth's atmosphere. 
The IACT technique was pioneered by the Whipple collaboration with the discovery of TeV emission from the Crab Nebula in 1989 \cite{Weekes1989}. After that, several IACT arrays (i.e. HEGRA\cite{Daum1997}, H.E.S.S.\cite{Konopelko2005},  MAGIC\cite{Alesic2016} and VERITAS\cite{Park2015}) allowed for a strong increase of the number of discovered sources in a few years period.

The success achieved by these instruments (all of them with a single mirror optics and cameras equipped with photomultiplier tubes) has driven the international VHE astronomy community towards the design of new-generation IACT telescopes, with the intent to enlarge the observable energy
region, to improve the sensitivity as well as to reach a wider field of view.
Double mirror optics and cameras with different sensors, equipped with a properly fast electronics, are the main objectives of such new designs.

In the field of new camera sensors, the Silicon Photo-Multipliers (SiPMs) represent the ideal choice offering all the advantages of a solid state device such as robustness  and low operative voltage. 
%Moreover, in recent decades, the nuclear physics, medical physics and astrophysics communities have given more impact to the partnership with its suppliers obtaining good improvements in the SiPMs performance. 
Several drawbacks have been considerably reduced, so obtaining the lowering of the breakdown voltage and of the optical cross-talk occurrences, a higher photon-detection-efficiency and a reduction almost twofold of the gain dependence on temperature variations. 
%It is precisely this latter feature that represents a major improvement to avoid significant variations in the pixels gain of the sensor. 
In addition, the performance of today's SiPMs makes it possible to observe in presence of a very high level of night sky background (NSB) without any damage  and loss of dynamic range induced by high level of current absorption.
Thanks to the previously described improvements, several new-generation IACT telescopes, first of all the small instrument FACT \cite{Anderhub2010},  have chosen SiPMs for their cameras. Among them, the ASTRI-Horn telescope whose experience concerning gain variations induced by high level NSB is described in the present paper.

The ASTRI-Horn telescope \cite{Lombardi2020} is an end-to-end prototype of the 4-m class IACTs developped by the Italian National Institute for Astrophysics (INAF)  in the context of ASTRI project (Astrofisica con Specchi a Tecnologia Replicante Italiana) \cite{Scuderi2018}. Its name has been chosen in honour of the Italian astronomer Guido Horn D'Arturo who pioneered the use of segmented primary mirrors in astronomy \citep{Horn1953}. ASTRI-Horn is equipped with UVscope \cite{Maccarone2021}, an auxiliary calibrated instrument, devoted to measure the NSB.  

Using data acquired in December 2018, during the commissioning phase of the telescope, it was possible to correlate the level of NSB measured by UVscope with the RMS of the pedestal for the data acquired by the ASTRI-Horn camera \cite{Compagnino2022}. Authors found that the pedestal RMS was not always linearly related to the NSB flux, as expected, but above a given level it becomes constant.  In particular, only when the NSB level detected by UVscope was lower than $\sim$ 4700 ph m$^{-2}$ ns$^{-1}$ sr$^{-1}$ (1.43 ph m$^{-2}$ ns$^{-1}$ deg$^{-2}$) the ASTRI-Horn camera response was nominal. The cause of the anomalous behaviour derived from a limit in the power supply that reduced the operative voltage of all its SiPM sensors. 

The aim of this paper is to demonstrate that the effects of this limitation are also present in the signals from the detected showers with a gain reduction. After some technical information on ASTRI-Horn telescope and UVscope (Sect.~\ref{sect:astrihorn}), the data reduction and analysis are presented in Sect.~\ref{sect:analysis}. Results and discussion are presented in Sect.~\ref{sect:results}; conclusions follow in Sect.~\ref{sect:discussion}.
\section{The ASTRI-Horn} 
\label{sect:astrihorn}

\begin{figure}[htb]
\begin{center}
\includegraphics[width=11.5cm] {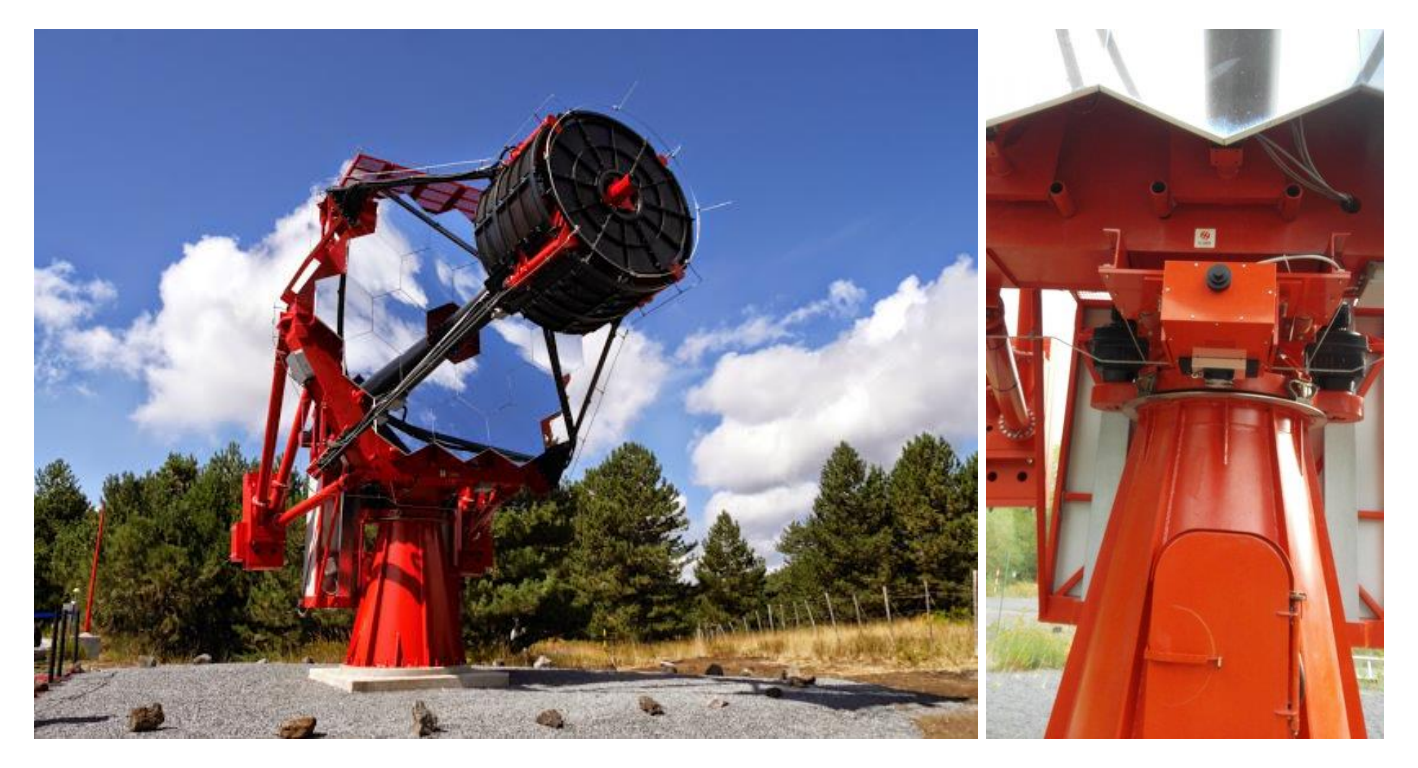}
\caption{ASTRI-Horn, the dual-mirror Cherenkov telescope installed on Mt.Etna, Italy, at the INAF 'M.C. Fracastoro' observing station (1740 m a.s.l.). On the rigth panel it is visible the small UVscope instrument mounted under the primary mirror structure of the ASTRI-Horn telescope.}
\label{fig:ASTRI_Horn}
\end{center}
\end{figure}

The ASTRI-Horn telescope (see Fig.\ref{fig:ASTRI_Horn}) is characterized by a dual-mirror optical system and a camera at the focal plane composed of multipixel SiPMs connected to a tailored fast (a few tens of ns) read-out electronics \cite{Sottile2016}. 
It is located on Mt. Etna, Serra La Nave, Italy, at the INAF 'M.C. Fracastoro' observing station \cite{Maccarone2013}. A complete description of the  telescope can be found in several papers and references therein (see e.g. \cite{Lombardi2020}). Here we focus on the telescope features that are of main interest for the analysis presented in Sect. \ref{sect:analysis}.

The telescope structure and its optics system have been chosen for the small-sized telescopes to be installed at the Cherenkov Telescope Array (CTA) \cite{Actis2011} southern site.  At the same time, the complete ASTRI-Horn is the prototype of the nine telescopes that form the ASTRI Mini-Array \cite{Antonelli2021} under construction at the Teide Observatory, Canary Islands, Spain, that will be able to study  
bright sources ($\sim$10$^{-12}$ erg cms$^{-2}$ s$^{-1}$ at 10 TeV)  with
an angular resolution of $\sim$3' and an energy resolution of $\sim$10\% at an energy of about 10 TeV \cite{Vercellone2022, Lombardi2021}.

The telescope optical system is based on a modified dual-mirror Schwarzschild-Couder configuration, \cite{Vassiliev2007} with the camera 
located between the secondary and primary mirror.
This camera is organized in 21 Photon Detection Modules (PDM), each one composed of 8$\times$8 multipixel SiPM sensors chosen for providing a compact system that reach a full field-of-view (FoV) of 7.6$^\circ$ \cite{Catalano2018}.The camera is thermally controlled to keep the temperature varying in a small interval ($\pm$2$^\circ$C) around the nominal value of 15$^\circ$C in order to control the gain stability at the level of 2\% \cite{Catalano2018}.

%\textbf{In contrast to other Cherenkov telescopes, that sample a digital waveform of the SiPM pulse}\cite{Sottile2016,Catalano2018},.....
This is accomplished thanks to the application-specific integrated circuit (ASIC) CITIROC (Cherenkov Imaging Telescope Integrated Read Out Chip) \cite{Fleury2014} properly customized for ASTRI-Horn.
The front-end electronics of the ASTRI-Horn camera is based on two separate chains that allow for high- and low-gain (HG and LG) amplification. During the period here examined the HG chain was not operative for scientific analysis, we considered only LG data.
The acquisition is based on a peak-detector technique, used for the first time in a Cherenkov camera. In contrast with other Cherenkov telescopes, that sample a digital waveform of the SiPM pulse, the peak-detector registers only the maximum of the shaped pulse height \cite{Catalano2018,Sottile2016}.
The read-out electronics is AC-coupled to the detector output blocking any slow varying signal as the NSB or star light in the FoV. However, Poisson fluctuations of these signals are detected by the electronics as fluctuations around the zero level (pedestal).
The standard deviation of these fluctuations depends on the NSB flux, the intrinsic electronic noise and the SiPM dark current. In operation, these fluctuations are mainly affected by NSB as the dark and electronic noise are much lower compared to NSB 
 \cite{Segreto2019}.

%{\color{red} D.I.: --- non mi risulta che sia stato corretto da Teresa ed Alessio --- T.M. ora è stata corretta
The ASTRI-Horn camera is also equipped with a fiber optics system (FOC) whose light, continuously or pulsed, illuminates the detector allowing to measure the gain uniformity over the focal plane with off-line analysis.
\\
A voltage distribution board (VDB) constituted by a main board and 21 daughter boards, one for each PDM, provides the required regulated power to all the electronics devices of the camera up to a current limit of  $\sim$6 mA \cite{Sottile2016}.

%Eventually, two light-tight lids prevent accidental sunlight exposure of the focal surface detectors \cite{Catalano2018}.\\
ASTRI-Horn implements a topological trigger in order to reduce background induced events. In particular the trigger is activated when a given number of contiguous pixels within a 
PDM measures a signal above a given threshold. Both the number of contiguous pixels required for the trigger and the signal threshold can be adjusted to minimize the number of false triggers.

\begin{figure*}[h!!]
\centering
\includegraphics[angle=-90, width=12.5cm] {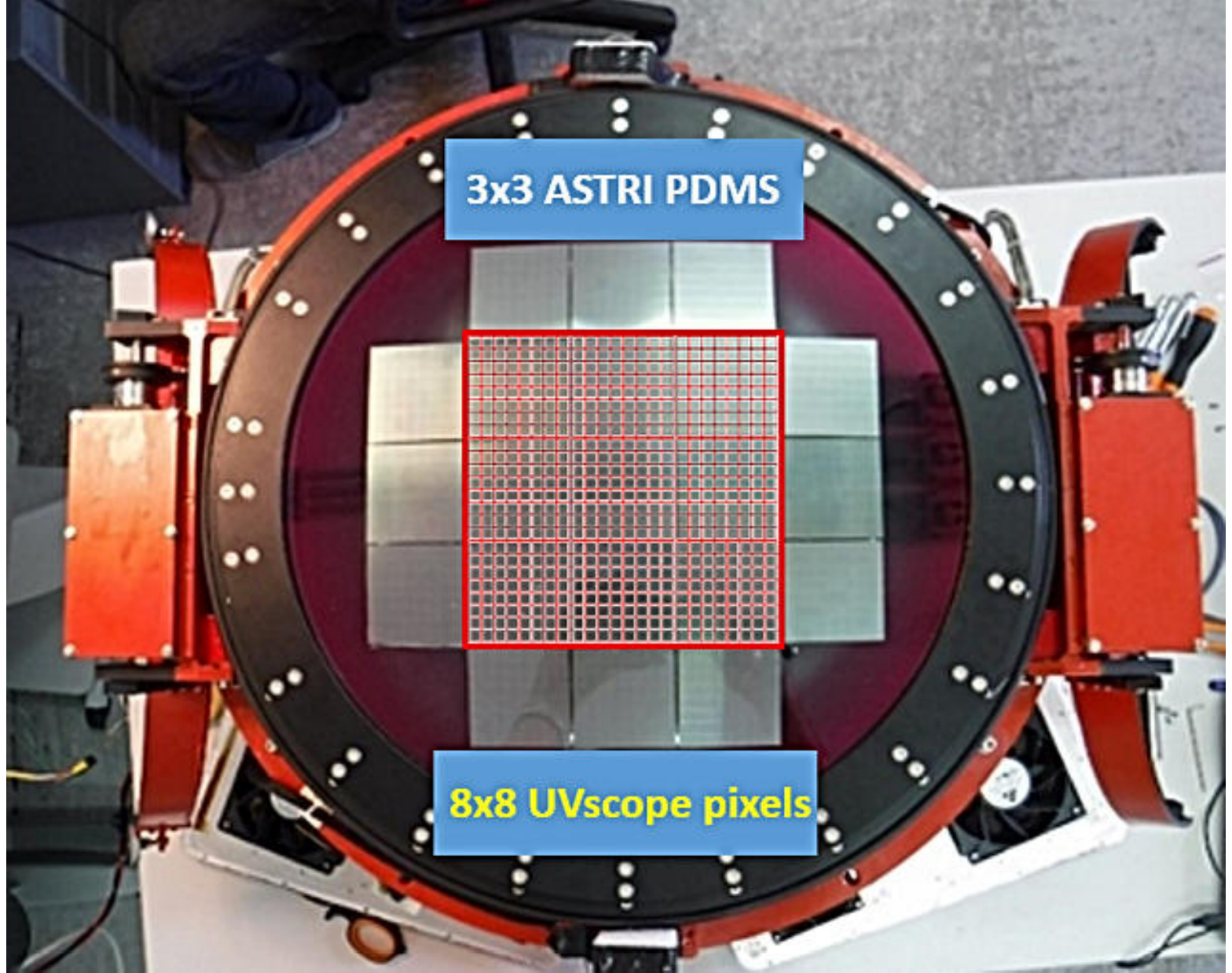}
\vspace{0.5cm}
\caption{Photo of the PDMs on ASTRI-Horn focal plane. The red square indicates the FoV seen by UVscope.}
\label{fig:focal_plane}
\end{figure*}

 ASTRI-Horn is equipped with UVscope, a light detector mainly devoted to measure the diffuse NSB in the wavelength band 300--650 nm \cite{Maccarone2021}. 
Mounted under the primary mirror structure of the telescope and co-aligned with the ASTRI-Horn camera axis, UVscope takes data contemporarily to and independently from the ASTRI-Horn camera acquisitions.
Its light sensor is a calibrated multi-anode photomultiplier tube (MAPMT) with 8$\times$8 pixels working in single photon counting mode \cite{Catalano2008}
to keep the electronic noise negligible. The angular aperture of each UVscope pixel (0.55') is equivalent to about 3$\times$3 pixels of the ASTRI-Horn camera. This makes the full UVscope FoV coincident with the one relative to the 3$\times$3 central PDMs of the ASTRI-Horn camera (identified in red colour in Fig.~\ref{fig:focal_plane}). 

The ASTRI-Horn scientific capability was proven by the detection of the Crab Nebula emission above an energy threshold of $\sim$3 TeV. The relative observations were carried out in December 2018 during the telescope commissioning phase for a total on-source observing time of 12.4 h, and an almost equal duration of the off-source exposure achieving a detection with a statistical significance of 5.4 $\sigma$ \cite{Lombardi2020}.  The same data also demonstrated the utility of UVscope in checking and quantifying the discrepancies with respect to the nominal working of the camera \cite{Compagnino2022}. In fact, the variance of the background  fluctuation in ASTRI-Horn camera is linearly correlated to the NSB flux if the camera is in nominal conditions. However, the analysis presented in \cite{Compagnino2022} showed that the linearity flattens at  NSB flux  levels higher than $\sim$4700 ph m$^{-2}$ ns$^{-1}$ sr$^{-1}$  when a current limit of $\sim$ 6mA in each PDM is reached.  

\section{Data reduction and analysis}
\label{sect:analysis}

The main task of our analysis is to check and quantify gain variations induced by high levels of the NSB in the acquired signals. To this purpose, we investigated how the distribution of the photoelectrons ($pe$) detected in the pixels of the recorded shower images varies in presence of high NSB fluxes.
Whenever relatively small gain variations (a few tens of percent) are present in the ASTRI-Horn camera, a given particle shower would produce a lower signal in the whole camera independently from its intensity without changing the shape of the $pe$ distribution. The mathematical model of this distribution, as directly derived from data, is a power law: 

\begin{equation}
N_r[pe, G_r]=p_0[G_r]\, pe^{p_1}
\end{equation}

\noindent
where  $N_r[pe,G_r]$  is the number of samples with $pe$ signal relatively to the nominal reference gain $G_r$; $p_0$ and $p_1$ are the normalization and the index of the power law, respectively. 
%Since $N_r[pe,G_r]$ depends on the run exposure,  any comparison between runs must be reported to the same exposure.
This formula for a different new gain value $G_n$ becomes:

\begin{equation}
N_n[pe,G_n]=p_0[G_r]\, (\frac{G_n}{G_r}\,pe)^{p_1}=p_0^{'}[G_n]\, pe^{p_1} 
%\label{Eq:pedistr2}
\label{eq:gain}
\end{equation}

\noindent
where $p_0^{'}[G_n]=p_0[G_r]\, (\frac{G_n}{G_r})^{p_1}$ is the new normalization of the power law.
This formula shows that it is the normalization parameter that follows any gain variations.

The same analysis can be performed in terms of digital signals  from the camera electronics (ADC count). In our case, for LG data, 1 $pe$ corresponds to about 1.5 ADC, and a few tens of percent in the gain variations is within one ADC.  Moreover, to enhance the statistics it is necessary to sum all the pixel distributions, that requires to align them because each pixel has a different pedestal.  For these reasons, we find more convenient to perform the analysis using $pe$ distributions, but we also checked, for one of the runs, that results are equivalent if ADC distributions are used.

\begin{figure*}[h!!]
\centering
\includegraphics[angle=-90, width=13.5cm] {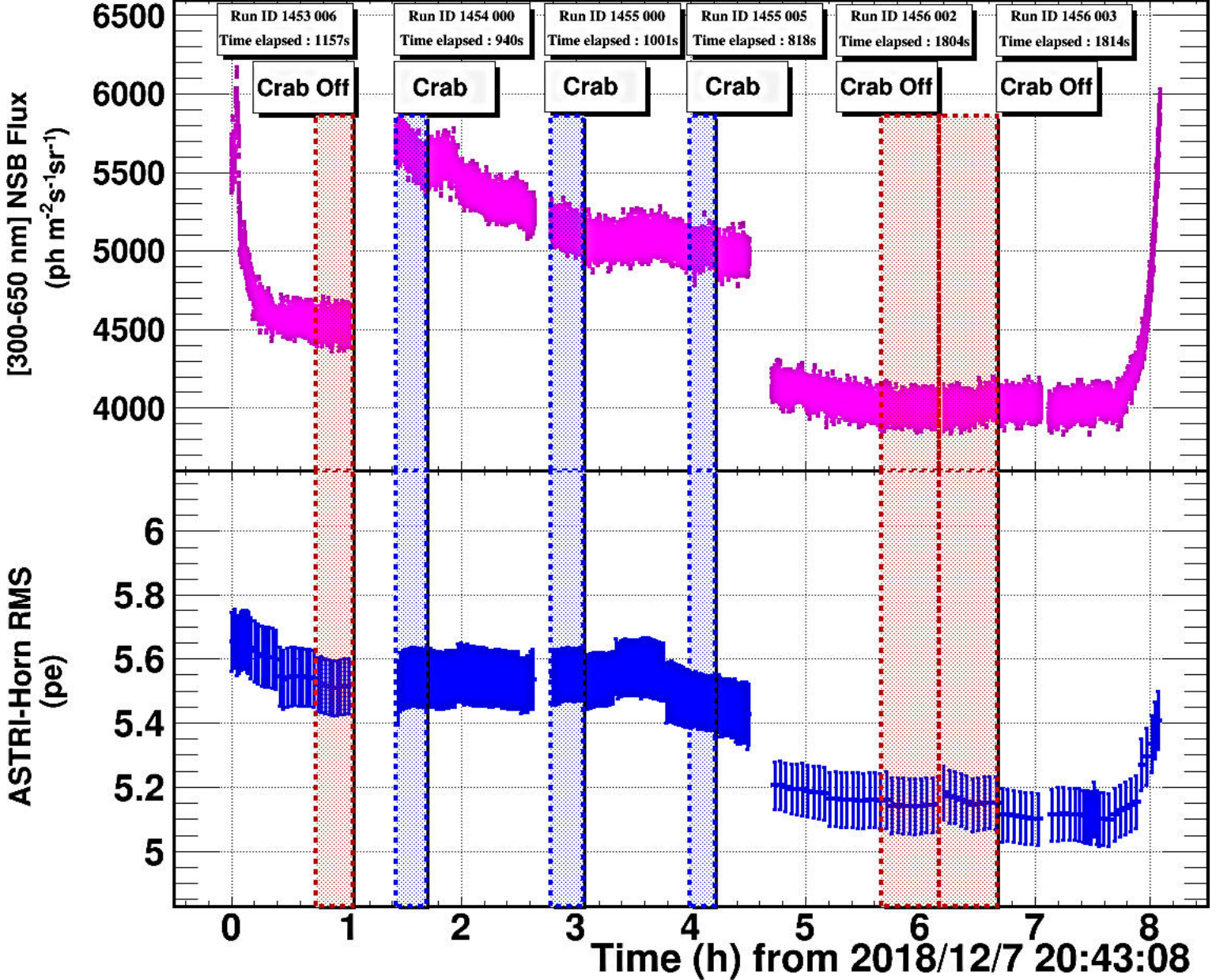}
\vspace{0.5cm}
\caption{Temporal evolution of the NSB flux measured by UVscope   (top panels) relative to the observing night 2018 December 7-8 \cite{Maccarone2021}; the bottom panel shows the contemporary evolution of the RMSs measured by ASTRI-Horn \cite{Compagnino2022}}. The vertical shaded areas identify the considered time intervals; the blue colour is relative to intervals affected by current limitation, while the red one shows the periods when the camera presents nominal behaviour.
\label{fig:nsb_time}
\end{figure*}

For our analysis, we  selected ASTRI-Horn  and UVscope contemporary acquisitions in six time intervals relative to different NSB levels during the 2018 December 7--8  observation night, where effects of the PDM current limitations are detected \cite{Compagnino2022}.
Table~\ref{tab:obslog} lists the selected Run IDs and the sequence numbers of the ASTRI-Horn considered files, together with the starting times, exposures, and pointing directions. The UVscope data, acquired along all the night towards the same ASTRI-Horn pointings, are stored in an unique single file. The plot of the temporal evolution of the NSB fux measured by UVscope every second during all nights is shown in the top panel of Fig.~\ref{fig:nsb_time}. 
The bottom panel of the figure shows, for comparison, the root mean squared ($RMS$) of background fluctuations  measured in the nine central PDMs of ASTRI-Horn camera, and averaged in $\sim$2 minutes \cite{Compagnino2022}. 
This $RMS$ was obtained as $\sigma$ of the comulative p.e. distribution around the pedestal after subtracting showers and star signals.
The three pointings Run ID 1453(006), 1456(002) and 1456(003), indicated with red shaded area in Fig.~\ref{fig:nsb_time}, were not affected by current limitation, at unlike the observations identified within the blue shaded area (Run ID 1454(000),1455(000) and 1455(005)).  The difference in the ASTRI-Horn response can be easily inferred by comparing the almost absence of variations in the $RMS$ between the first run (ID 1453(006); T= 0.32 h) and the second one (ID 1454(000); T= 0.26 h) with respect to UVscope data whose intensity increases by $\sim$25\% \cite{Compagnino2022}. 

\begin{figure*}[ht]
\centering
\includegraphics[angle=-90, width=13.5cm] {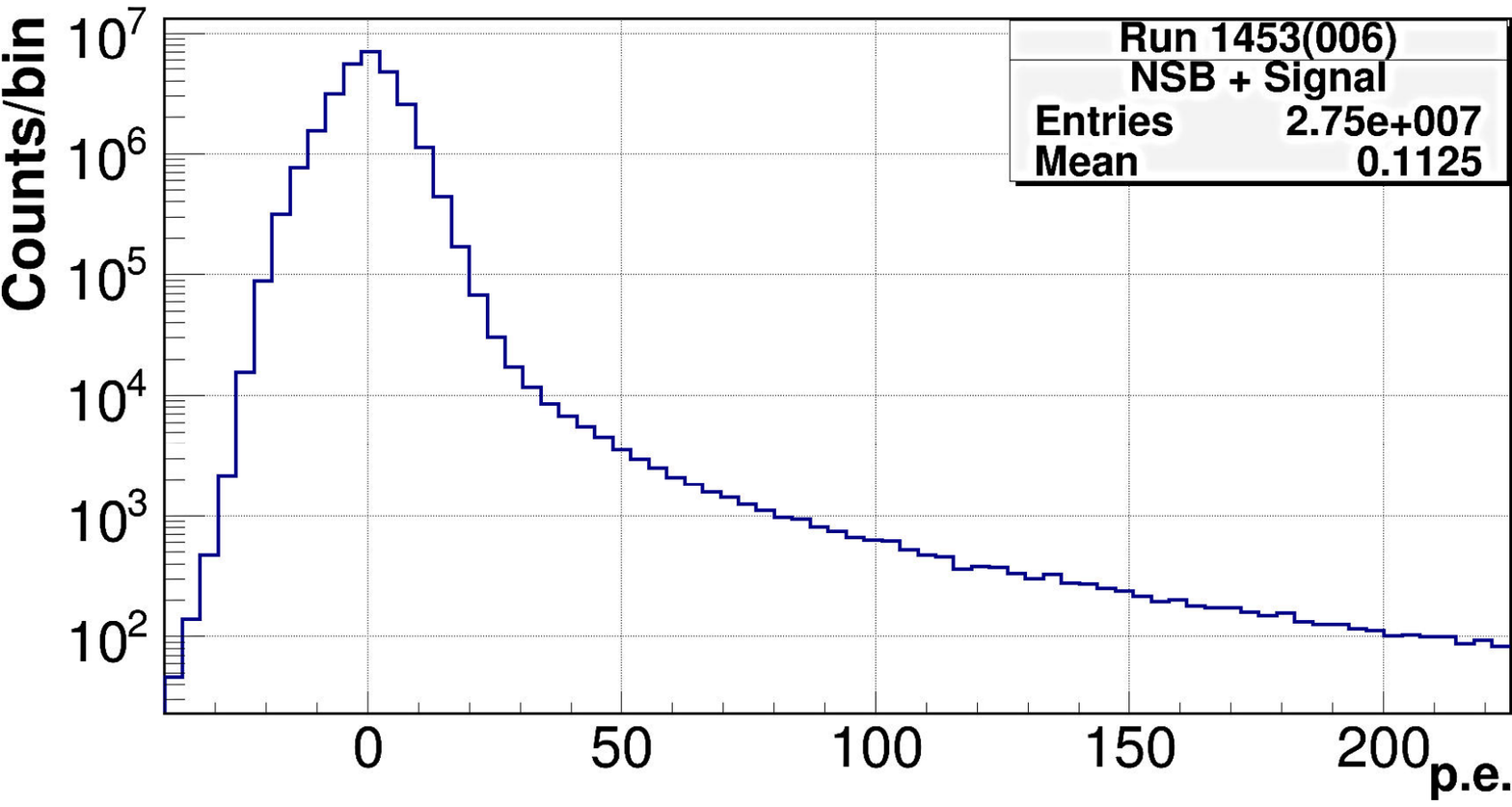}
%\vspace{0.5cm}
\caption{Run 1453(006) $pe$ distribution of all the signals produced by both the night sky background and  shower events,  in the nine central PDMs.}
\label{fig:pe_distribution_all_signal}
\end{figure*}

\begin{figure*}[ht!!]
\centering
\includegraphics[angle=-90,width=15.5cm] {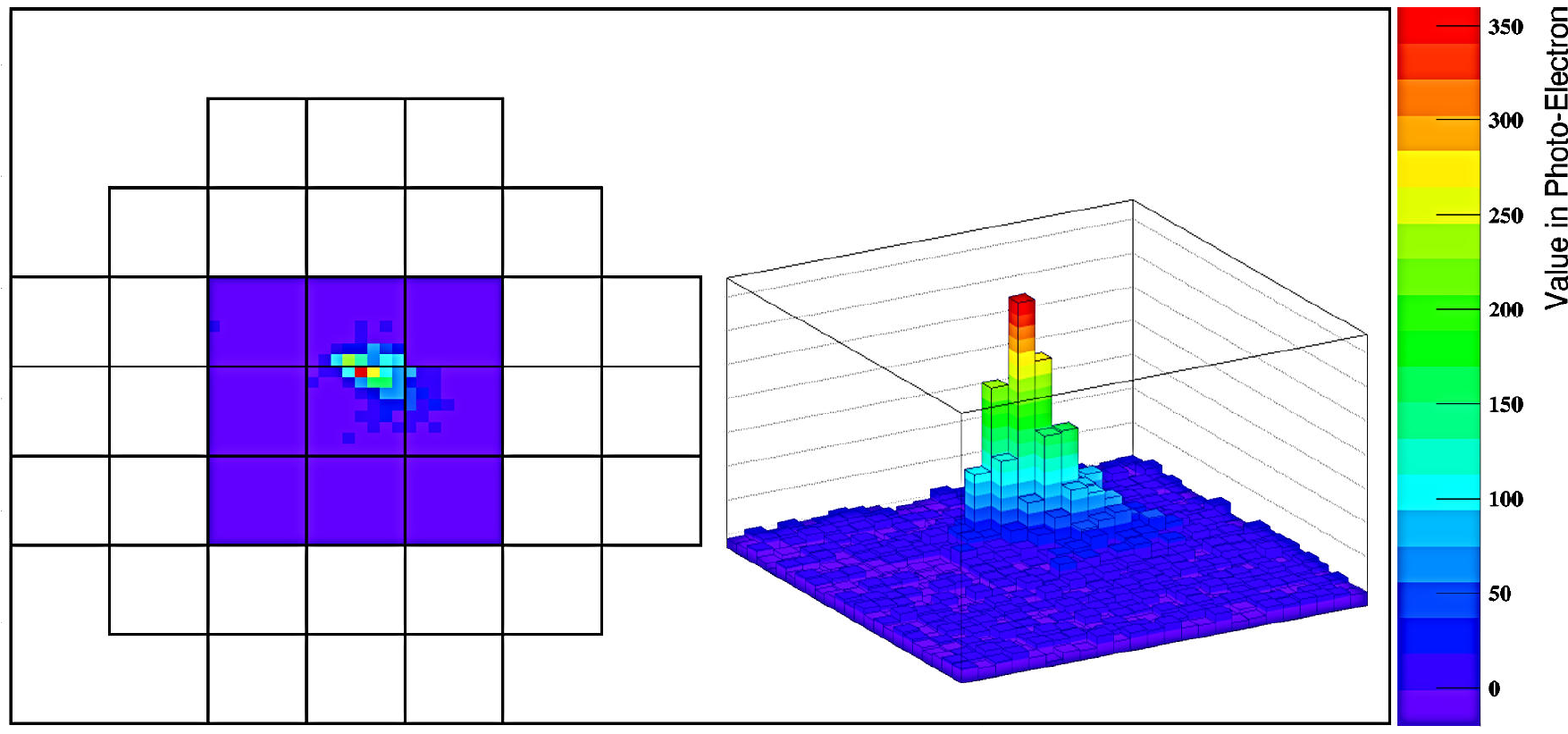}
\caption{Shower event in the Run 1453(006) where the NSB and the signal pixels can be observed.} 
\label{fig:pe_image}
\end{figure*}

Only the nine central PDMs (see Fig.~\ref{fig:focal_plane}) were considered in the analysis presented in this paper allowing for a comparison with the results presented in  \cite{Compagnino2022}. %Data were relative to the LG chain, being the HG not available at the considered observing time.
The first step of the data reduction is the conversion
%from ADC counts to $pe$ with the method explained in detail in \cite{Compagnino2022}.  
%and with the data listed in Table~2 and 3 of the same paper . It concerns the relative calibration of the pixels gain in the camera, i.e. the signal extraction and conversion into physical units, and the estimation of the level of statistical fluctuations of the pixel signals. Using an optical fiber calibration system (FOC) inside the ASTRI-Horn camera, kept closed lid, it was possible to acquire signal waveforms with a sampling in peak detector mode, in order to obtain a pulse height distributions (PHD) for each physical pixel. However the calibrated data must be expressed in term of photo-electrons(p.e.). Therefore the conversion factors from the maximum amplitude of the pulse to p.e. and from the integral of the pulse to p.e. must be known. Each peak in the spectrum corresponds to a certain number of photo-electrons. The first peak (0 p.e.) contains all the events where no photon was detected and is referred to as "pedestal" peak. The second peak (1 p.e.) corresponds to the signal of one detected photon, and so forth. The clear separation of the individual peaks shows the excellent single photon resolution which can be achieved due a high uniformity in the response of the individual micro-cells of each SiPM pixel. The  peaks are in good approximation Gaussian shaped. The distance between two neighboring peaks corresponds to the gain of each pixel since it indicates the amount of charge produced in an avalanche breakdown of one SiPM pixel.}
of analog signals coming from the camera electronics, digitized through external ADC devices, to $pe$.  The following relation was used:
\begin{equation}
{
pe=\frac{ADCcounts-Pedestal}{Gain} 
}
\end{equation}    

\noindent
where $Gain$ and $Pedestal$, for each pixel, are determined from dedicated calibration measurements as explained in detail in \cite{Compagnino2022}.
A single $pe$ distribution was accumulated for each run; in this case,  $N_r[pe,G_r]$ in each bin depends on the run exposure, any comparison between runs must take into account this. Fig.~\ref{fig:pe_distribution_all_signal} shows, as an example, the distribution relative to the Run 1453(006). The distribution curve presents two components: a Gaussian around the pedestal due to the NSB and used for the analysis in \cite{Compagnino2022}, and a power law produced by the shower signals. 
Fig.~\ref{fig:pe_image} shows a shower event in Run 1453(006) where the NSB and the signal pixels can be observed.

\begin{figure*}[ht]
\centering
\includegraphics[angle=-90, width=13.5cm] {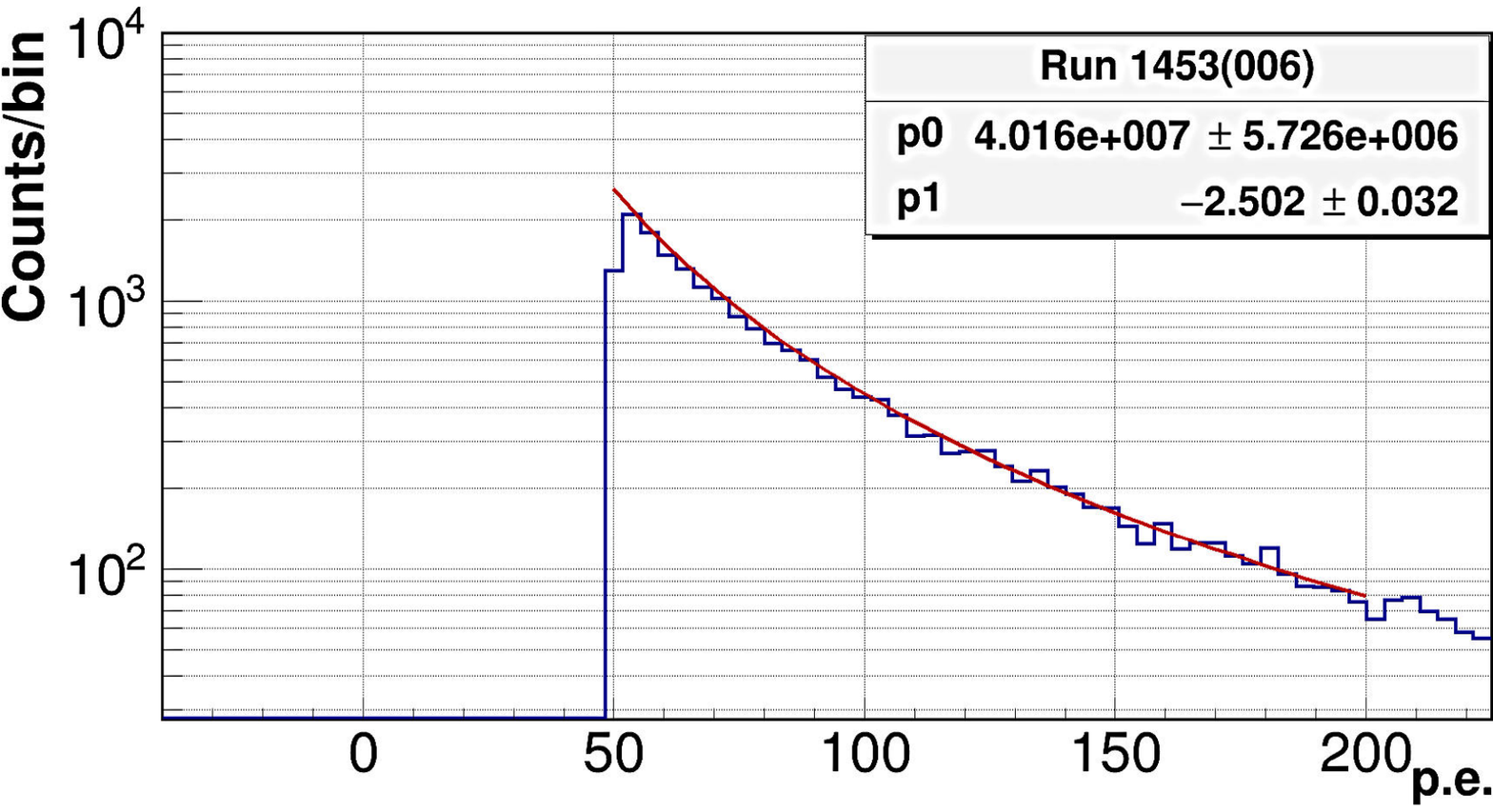}
%\vspace{0.5cm}
\caption{Run 1453(006) p.e. distribution of the images detected in the nine central PDMs. The red curve is the correspondent best fit power law.}
\label{fig:pe_distribution}
\end{figure*}

Since we are interested in pixels whose signal is produced by showers and whose content is well above the NSB, only pixels with more than 50 $pe$ were considered to accumulate the distributions.  The final curves include only the power law component, as shown in Fig.~\ref{fig:pe_distribution} for the Run 1453(006), as an example. 
%Whever the NSB increases, an increase of the  Gaussian Full Width at Half Maximum is produced. } 

\begin{table}[ht]
\caption{ Observation log of ASTRI-Horn data used in this paper}.
\centering
\begin{tabular}{cccl}
\hline\hline
\multicolumn{1}{c}{ Run ID (Seq.)} & 
 UTC Starting Date      &
\multicolumn{1}{c}{Exposure}& \multicolumn{1}{c}{Pointing} \\
   & (d h:m:s)   & \multicolumn{1}{c}{(s)}   &\multicolumn{1}{c}{RA,Dec}  \\
\hline     
1453 (006)  & 2018/12/07 21:27:10 &  1157 & 46.13$^\circ$, 22.01$^\circ$ (Crab Off)  \\ 
1454 (000)  & 2018/12/07 22:09:16 &  940  &   83.63$^\circ$,   22.01$^\circ$ (Crab)  \\ 
1455 (000)  & 2018/12/07 23:30:08 &  1001 &  83.63$^\circ$,    22.01$^\circ$ (Crab)  \\ 
1455 (005)  & 2018/12/08 00:42:25 &  818  &  83.63$^\circ$,     22.01$^\circ$ (Crab)  \\ 
1456 (002)  & 2018/12/08 02:23:00 &  1804 &  121.13$^\circ$, 22.01$^\circ$ (Crab Off) \\
1456 (003)  & 2018/12/08 02:53:04 &  1814 & 121.13$^\circ$, 22.01$^\circ$ (Crab Off)\\ 
\hline\hline
\end{tabular}
\label{tab:obslog}
\end{table}

%We will present and discuss this in Sect.~\ref{sect:results}. 

\newpage

\subsection{Analysis of the $pe$ distributions}
%The p.e. distributions from the central PDMs were accumulated for each run; as an example the distributions relative to Run 1453(006) is shown in Fig.~\ref{fig:pe_distribution}. 

The distribution accumulated for each Run  was then modelled with the power law function of Eq.\ref{eq:gain}. 
The two parameters of the model were derived with a fitting procedure performed in the interval 50--200 $pe$ where each bin has more than 20 counts, that allows for assuming a Gaussian statistic and  using the $\chi^2$ test to estimate the goodness of the fit. Errors in the fitting parameters are relative to 90\% confidence level.
%chosen to exclude from the analysis those pixels where still the contribution from faint stars or muons could be not negligible. 

%%These two parameters were derived with a fitting procedure performed in the %%interval 50--200 p.e., chosen to exclude from the analysis those pixels where %%the contribution from the background,  faint stars or muons could be not %%negligible and to guarantee the applicability of the $\chi^2$ statistics %%having all bins of the six distributions more than 20 samplings.

The final parameters were obtained in two steps with an iterative procedure.  All distributions were, at first, fitted leaving $p_{0}$ and $p_{1}$ as free parameters and latter all $p_1$ were fixed to the average obtained from the values derived in the first step.  This procedure avoids the effects of the correlation between $p_0$ and $p_1$ in the fitting procedure. 
All runs present different temporal duration, then to compare the $p_0$ value of each run we normalized them to the correspondent exposure.

\begin{figure*}[ht!!]
\centering
\includegraphics[angle=-90,width=15.5cm] {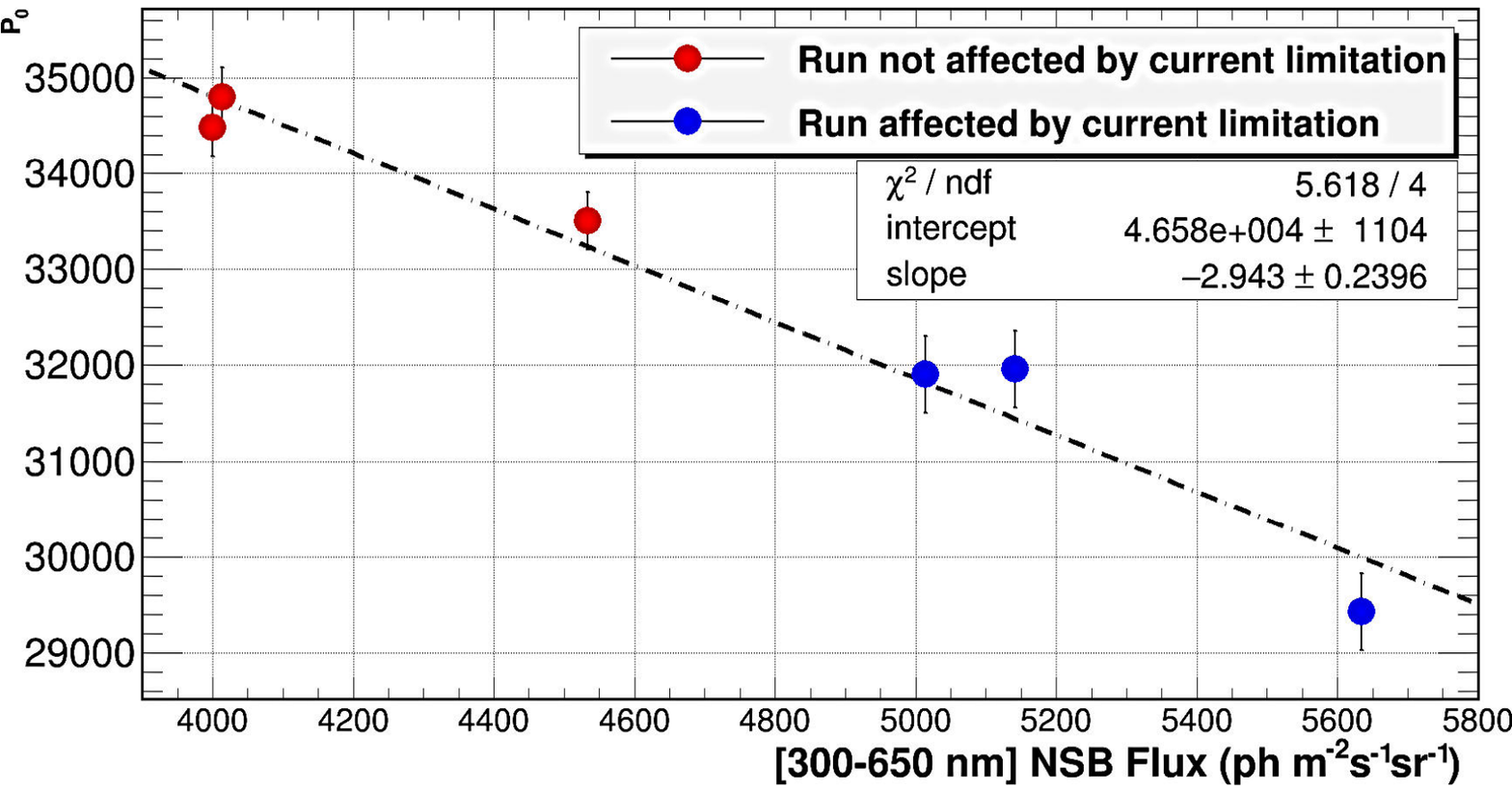}
\caption{Best fit $p_0$ values vs  NSB fluxes measured by UVscope. The black dash-dotted line of the best fit shows the  linear trend as the flux increases.} 
\label{fig:pe_norm}
\end{figure*}

\begin{figure*}[ht]
\centering
\includegraphics[angle=-90, width=13.5cm] {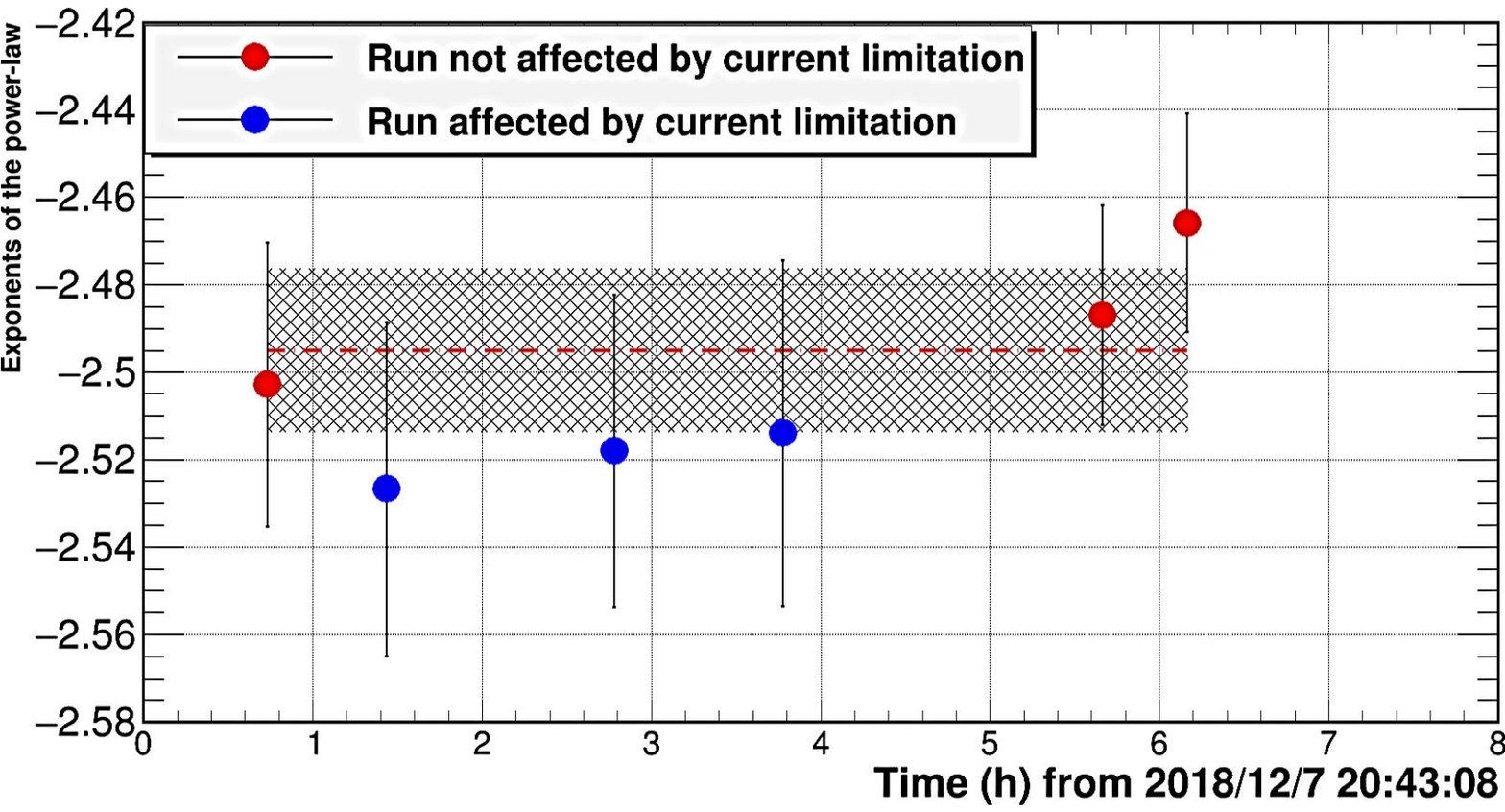}
%\vspace{0.5cm}
\caption{Best fit value of the power law indices vs time obtained from the first step of the fitting procedure leaving both $p_0$ and $p_1$ as free parameters. The red and blue points indicate the runs with a lower and higher NSB, respectively.
The best fit of  all points is represented with a dashed red line and its  error with a shaded area. 
}

\label{fig:index}
\end{figure*}

\begin{figure*}[ht]
\centering
\includegraphics[angle=-90, width=13.5cm] {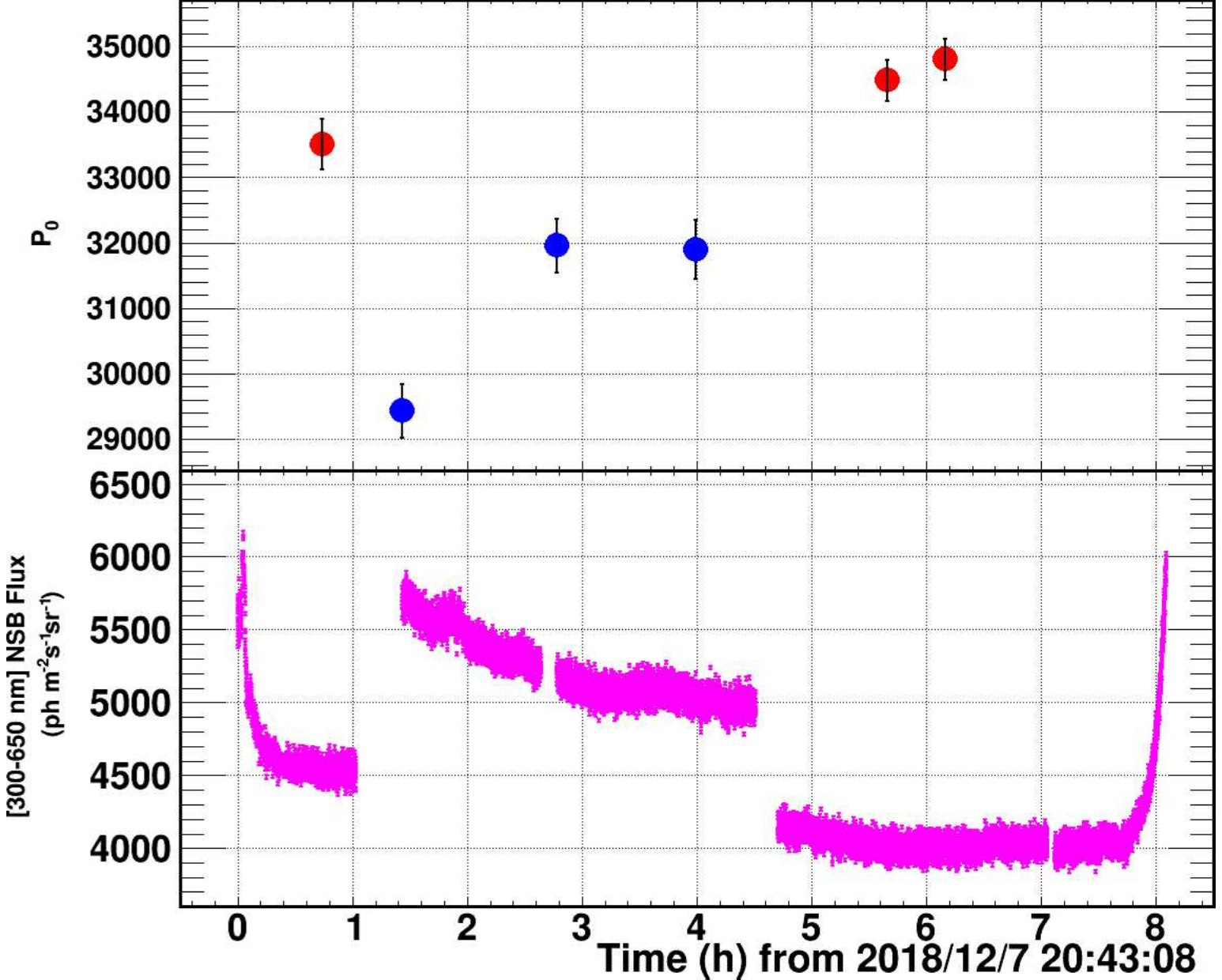}
%\vspace{0.5cm}
\caption{Best fit $p_0$ values, normalized for the exposures, obtained with fixed indices is shown in the top pane as function of the time:  red points are relative to runs not affected by gain reduction, while the blue ones mark intervals with nominal functioning. Errors are relative to 90\% confidence level.
The contemporary NSB flux measured by UVscope is shown in the bottom
panel. }
\label{fig:normalization}
\end{figure*}

The best fit parameters relative to the first step are shown in Table~\ref{tab:p0p1} together with the reduced $\chi^2$, for the six analysed runs. In all cases, the adopted model reproduces well the distributions being all $\chi^2$ within 2$\sigma$ of the expected values.   
Indices \textbf{$p_1$} are also shown in Fig.~\ref{fig:index} as function of time.
Fitting all values with a constant we obtained 2.49$\pm$0.02 used in the second step of the fitting procedure.
In Fig.~\ref{fig:index} this constant is represented with a dashed red line and its  error with a shaded area.
The $p_0$ parameters obtained in the second step are presented in Fig.~\ref{fig:normalization} and in Table~\ref{tab:p0}, together with the values of the reduced $\chi^2$  that state again the goodness of the model. In the same table, the average NSB fluxes measured by UVscope are also shown.
The Fig.~\ref{fig:pe_norm} shows the best fit line of  $p_0$ values with a slope of 2.9 and an intercept of 4.6$\cdot10^{4}$.
The $\chi^2$ distribution provides an acceptable match within the 2 $\sigma$. 
The behavior of $p_0$ points is linear, since they are variations of the order of several tens in percentage.
The linear trend is decreasing with the increased NSB flux measured by UVscope.
\newpage
\begin{table}[ht]
\caption{Fitting results for the analysed distributions using a power law model with all parameters free. The reduced $\chi^2$ are relative to 48 dof. }
\centering
\begin{tabular}{cccc}
\hline\hline
\multicolumn{1}{c}{ Run ID (Seq.)} & $p_0$ & $p_1$ & $\chi^2$ \\
                                   & 10$^4$ Counts/bin/s  &  & \\
\hline     
1453 (006)  & 3.47$\pm$0.49 & -2.50$\pm$0.03  &  0.91 \\ 
1454 (000)  & 3.33$\pm$0.55 & -2.52$\pm$0.03   & 1.03   \\ 
1455 (000)  & 3.51$\pm$0.55 & -2.51$\pm$0.03   & 0.90 \\ 
1455 (005)  & 3.43$\pm$0.59 & -2.51$\pm$0.04   & 1.33   \\ 
1456 (002)  & 3.33$\pm$0.36 & -2.48$\pm$0.02   & 0.92  \\
1456 (003)  & 3.07$\pm$0.33 & -2.46$\pm$0.02   & 0.94\\ 
\hline\hline
\end{tabular}
\label{tab:p0p1}
\end{table}

\newpage
\begin{table}[ht]
\caption{Best fit value of the $p_0$ parameters obtained from the fit with fixed index (49 dof). The UVscope fluxes averaged over the considered intervals  are also reported for comparison. Errors in UVscope fluxes are the $RMS$ respect to the mean.}

\centering
\begin{tabular}{cccc}
\hline\hline
\multicolumn{1}{c}{ Run ID (Seq.)} & $p_0$ & $\chi^2$  & 300--650 NSB Flux  \\
 & 10$^4$ Counts/bin/s  &   &\multicolumn{1}{c}{ph m$^{-2}$ ns$^{-1}$ sr$^{-1}$} \\
\hline     
1453 (006)  & 3.35$\pm$0.03 &  0.89  & 4534$\pm$56  \\ 
1454 (000)  & 2.94$\pm$0.04 &  1.05  & 5635$\pm$84   \\ 
1455 (000)  & 3.20$\pm$0.04 &  0.90  & 5141$\pm$63   \\ 
1455 (005)  & 3.19$\pm$0.04 &  1.32  & 5014$\pm$59    \\ 
1456 (002)  & 3.45$\pm$0.03 &  0.90  & 4000$\pm$52  \\
1456 (003)  & 3.48$\pm$0.03 &  1.00  & 4014$\pm$53   \\ 
\hline\hline
\end{tabular}
\label{tab:p0}
\end{table}

\newpage
\section{Results and Discussion}
\label{sect:results}

The $pe$ distributions used in our analysis, after the exposure normalization, should not have been correlated to the NSB levels if the camera had operated in nominal regime.
However, results presented in Fig.~\ref{fig:normalization} show that all time intervals non affected by current limitation have compatible normalization values if a 2\% calibration systematic is also included, while significant decreases are detected in the interval with high NSB levels (time 1$^h$:30' to 4$^h$:30' from the observation starting time). 
The maximum decrease is observed in correspondence of the highest level of the diffuse NSB. To quantify this decrease, we computed the ratio between the values relative to runs affected by current limitation  and the average between the two ones correspondent to the lowest NSB level (1456(002) and 1456(003)). The highest value, relative to run 1454(000), is $\sim$15\%.

\begin{figure*}[h!!]
\centering
\includegraphics[angle=-90, width=13.5cm] {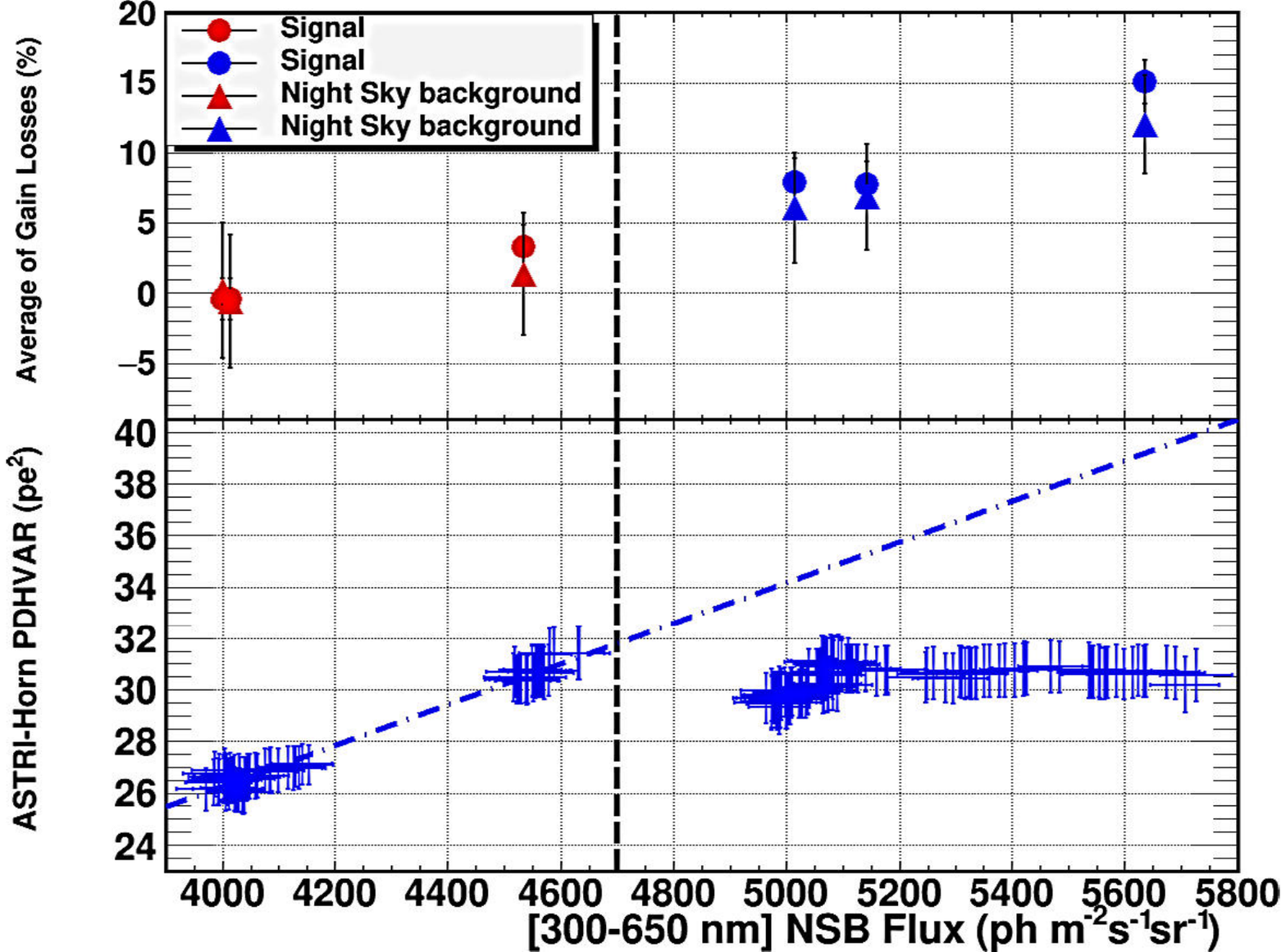}
\vspace{0.5cm}
\caption{The top panel shows the gain losses, in percentage, computed from the $PHDVAR$ (dots) and from the fast signals (triangles) as function of the  UVscope fluxes.
The bottom panel shows the ASTRI-Horn $PHDVAR$s vs 300-650 nm NSB fluxes ($PHDVAR$ = RMS$^{2}$, see \cite{Compagnino2022}). The blue dash-dotted line indicates the best fit in the nominal working regime. The dashed vertical line indicates the level above which the camera is affected by current limitation.}
\label{fig:gainvar}
\end{figure*}

A detailed comparison of the results obtained both from the analysis of atmospheric showers signals and those from the NSB \cite{Compagnino2022}, is necessary to assess that the current limitation affects both at the same level.
To this purpose, we evaluated the percentage gain variations from the nominal values in both data sets:\\
{\textit{Signal}}:  in this case, the  normalization $p_0[G_r]$, was obtained averaging the values relative to the lowest NSB (runs 1456(002) and 1456(003)) to improve the significance. The percentage variations was then computed with respect to this value for all other runs.
The gain variation  $\Delta G[NSB]$  is given by:
\begin{equation}
\Delta G[NSB]=1-\frac{p_0[NSB]}{p_0[G_r]}
\label{Eq:pedistr3}
\end{equation}
where $p_0[NSB]$ is the normalization relative to each run.
Results are shown  in the top panel of Fig.~\ref{fig:gainvar} with dots.
\\
{\textit{NSB}}:
in this data set, the evaluation of the NSB statistical variance for each pixel should has been based on the $Variance$ data obtained by the camera electronics \cite{Segreto2019}. Unfortunately these data were not available in December 2018 due to a technical problem.  To overcome such absence, the NSB variance, hereafter named $PDHVAR$, has been computed as the width of the $pe$ distribution around the pedestal fitted with a Gaussian, considering that the contribution due to the intrinsic electronic noise is negligible. \cite{Compagnino2022}.
%\textcolor{blue}{\textbf{$PDHVAR$ represents the sum of the intrinsic standard deviation of the electronics plus the SiPM noise observed in dark conditions ($PDHVAR_{dark}$), and the $PHDVAR_{NSB}$ which is the standard deviation induced by the NSB level as described by the following formula $PHDVAR = PHDVAR_{dark} + PHDVAR_{NSB}$ in \cite{Compagnino2022}.}}
Note that a check when $Variance$ data were present, as in March 2019, confirmed that the two values are equivalent within a systematic error of 1\% \cite{Compagnino2022}.
For our analysis, we fitted the $PDHVAR$s in the nominal region as function of the NSB flux measured by UVscope  and  extrapolated this line to all the interval of NSB flux we are considering (dashed blue line in Fig.~\ref{fig:gainvar}). We then fitted the $PDHVAR$s in the intervals relative to  runs affected by current limitation with a constant  and computed the ratios with respect to the best fit line. Using a formula analogue Eq.\ref{Eq:pedistr3} we computed the gain variation for the NSB.
No gain variation is detected  
below 4700 ph m$^{-2}$ ns$^{-1}$ sr$^{-1}$. Resulting values in percentage are shown in Fig.~\ref{fig:gainvar} with  triangles. 

It is well evident from Fig.~\ref{fig:gainvar}, that the  gain
variations detected with the two analyses are compatible within errors. We can then conclude that the phenomenon, induced by the high level of NSB, affects equally the background and the shower signals.
% il primo punto ha una variazione compatibile con la variazione del gain sulla camera 2\%

In all the data available from the 2018 Crab observing campaign, the gain reduction affects only the observations with the telescope pointing at the Crab Nebula where a high level of NSB is present (see Fig.~\ref{fig:nsb_time}). The gain correction presented in this paper therefore would lead, if applied, to an improvement of the already important result  of the Crab detection statistical significance (5.4$\sigma$) published in \cite{Lombardi2020}. 
%where the number of background events from the OFF-source intervals was corrected only for the different observing time (factor $\alpha$=0.88).} 
\\

\section{Summary and Conclusions}
\label{sect:discussion}
In this paper, we presented results on the gain variation that affected the shower signal detected by ASTRI-Horn as consequence of  the increase of the NSB flux, even if the gain calibration of each camera pixel performed periodically presented a stable behaviour at level of 2\% along all the 2018-2019 observations.
The analysis presented in this paper was performed using a set of data collected on December 2018. We found that only data collected during Crab pointings were affected by gain reduction with a maximum  of $\sim$15\% observed in correspondence of the highest NSB level of 5635 ph m$^{-2}$ ns$^{-1}$ sr$^{-1}$(1.71 m$^{-2}$ ns$^{-1}$ deg$^{-2}$). Off-source data,  on the contrary, are relative to a nominal functioning of the camera.
%being the NSB level always lower than 4700 ph m$^{-2}$ ns$^{-1}$ sr$^{-1}$.  
Moreover, the levels of gain variation measured with our current analysis is compatible within the errors  with those obtained from the NSB analysis as presented in our previous work \cite{Compagnino2022}.

The limit in the PDM power system of ASTRI-Horn has been upgraded to reduce this problem. The new  power system will be able to supply a constant power up to $\sim$14 mA that will allows us for a correct observation in dark time (Moon illumination $<$40\%). Furthermore, this limit has been moved to $\sim$70 mA for the telescopes of ASTRI Mini-Array to make them able to observe also in grey conditions (Moon illumination between 40\% and 70\% and Moon minimum angular distance $>$90$^\circ$).

We highlight the importance of the method presented in this paper that will be used in the new  ASTRI-Horn and ASTRI Mini-Array observation campaign, to select the "good time interval" (no clouds, nominal functioning, low NSB level).

The main point that this paper highlighted is the importance of having a contemporary measurements of the level of NSB, not only for a better modelling the background level in  shower  images, but also as diagnostic tool for the correct functioning of the telescope. 

To this purpose, UVSiPM, an upgraded version of UVscope is foreseen for the ASTRI Mini-Array. This  will use the same sensors and the same filter as the telescope cameras for simpler correlation between the two instruments. 
In extreme observing conditions, as with high Moon illumination, the use of this auxiliary instrument would be very useful in supporting the scientific analysis.

\acknowledgments
Authors thank the anonymous referee for his/her improving the clarity of the paper and Saverio Lombardi for his insightful comments.
This work was conducted in the context of the ASTRI Project, supported by the Italian Ministry of Education, University, and Research (MIUR) with funds specifically assigned to the Italian National Institute of Astrophysics (INAF), and by the Italian Ministry of Economic Development (MISE) within the Astronomia Industriale program. This work has gone through internal review by the ASTRI Project Collaboration.
%%\end{acknowledgements}

\noindent
{\bf Data Availability Statement}:
The data  analysed during the current study are not publicly available being taken during  the commissioning phase of the telescope,  but are available from the corresponding author on reasonable request.

\noindent
{\bf Conflicts of Interest}:
The authors declare that they have no conflicts of interest.
\bibliographystyle{JHEP}

\providecommand{\href}[2]{#2}\begingroup\raggedright\endgroup

\end{document}